\documentclass[useAMS,usenatbib]{mn2e} 
\usepackage{bm}
\usepackage{graphicx}

\def\lesssim{\mathrel{\hbox{\rlap{\hbox{\lower4pt\hbox{$\sim$}}}\hbox{$<$}}}}
\def\gtrsim{\mathrel{\hbox{\rlap{\hbox{\lower4pt\hbox{$\sim$}}}\hbox{$>$}}}}
\newcommand{\mincir}{\raise
-2.truept\hbox{\rlap{\hbox{$\sim$}}\raise5.truept
\hbox{$<$}\ }}
\newcommand{\magcir}{\raise
-2.truept\hbox{\rlap{\hbox{$\sim$}}\raise5.truept
\hbox{$>$}\ }}
\newcommand{\siml}{\raise -2.truept\hbox{\rlap{\hbox{$\sim$}}\raise5.truept
\hbox{$<$}\ }}
\newcommand{\simg}{\raise -2.truept\hbox{\rlap{\hbox{$\sim$}}\raise5.truept
\hbox{$>$}\ }}
\newcommand{\be}{\begin{equation}}
\newcommand{\ee}{\end{equation}}
\newcommand{\ba}{\begin{eqnarray}}
\newcommand{\ea}{\end{eqnarray}}

\title[Shapley Optical Survey II]{Shapley Optical Survey II: The effect of environment on the colour-magnitude
  relation and galaxy colours\thanks{Based on European Southern
  Observatory archive data}}

\author[C. P. Haines et al.]{C. P. Haines,$^{1}$\footnotemark[7]
P. Merluzzi,$^1$ A. Mercurio,$^1$ A. Gargiulo,$^1$ 
N. Krusanova,$^1$ 
\newauthor
G. Busarello,$^1$ F. La Barbera,$^1$ and M. Capaccioli$^{1,2}$
\\
$^{1}$INAF - Osservatorio Astronomico di Capodimonte,
via Moiariello 16, I-80131 Napoli, Italy\\
$^{2}$Department of Physics, Universit\`{a} ``Federico II'', Napoli, Italy}

\begin{document}
\date{Accepted 1988 December 15. Received 1988 December 14; in original form 1988 October 11}

\pagerange{\pageref{firstpage}--\pageref{lastpage}} \pubyear{2002}

\maketitle

\label{firstpage}

\begin{abstract}
We present an analysis of the effects of environment on the
photometric properties of galaxies in the core of the Shapley
Supercluster at $z=0.05$, one of the most massive structures in the local
universe. The Shapley Optical Survey (SOS) comprises archive WFI
optical imaging of a 2.0\,deg$^2$ region containing the rich clusters A3556, A3558 and
A3562 which demonstrate a highly complex dynamical situation
including ongoing cluster mergers.
The \mbox{$B-R/R$} colour-magnitude relation has an intrinsic dispersion of
0.045\,mag and is \mbox{$0.015\pm0.005$}\,mag redder in the highest-density regions,
indicative of the red sequence galaxy population being 500\,Myr older
in the cluster cores than towards the virial radius. 
The $B-R$ colours of galaxies are dependent on
their environment, whereas their luminosities are independent of the
local density, except for the very brightest galaxies
\mbox{(M$_{R}\!<\!-22$)}. 
The global colours of faint \mbox{($\ga\!{\rm M}^{*}+2$)} galaxies change from the
cluster cores where \mbox{$\sim\!90$\%} of galaxies lie along the cluster
  red sequence to the virial radius, where the fraction has dropped to just
  \mbox{$\sim\!20$\%.}
This suggests that processes related to the supercluster
environment are responsible for
transforming faint galaxies, rather than galaxy merging, which should
be infrequent in any of the regions studied here.
The largest concentrations of
faint blue galaxies are found {\em between} the clusters, coincident
with regions containing high fractions of \mbox{$\sim\!L^{*}$} galaxies with
radio emission indicating starbursts. Their location suggests
star-formation triggered by cluster mergers, in particular the merger of
A3562 and the poor cluster SC\,1329-313, although they may also
represent recent arrivals in the supercluster core complex. 
The effect of the A3562-SC\,1329-313 merger is also apparent as a
displacement in the spatial distribution of the faint galaxy population
from both the centres of X-ray emission and the brightest cluster
galaxies for both systems. The cores of each of the clusters/groups
are marked by regions that have the lowest blue galaxy fractions and
reddest mean galaxy colours over the whole supercluster region,
confirming that star-formation rates are lowest in the cluster
cores. In the cases of A3562 and SC\,1329-313, these regions
coincide with the centres of X-ray emission rather than
the peaks in the local surface density, indicating that ram-pressure
stripping may have an important role in terminating any remnant
star-formation in galaxies that encounter the dense ICM of the cluster cores.

\end{abstract}


\footnotetext[7]{E-mail:chris@na.astro.it}
\section{Introduction}
\label{intro}

The cluster galaxy population has evolved rapidly over the last 4\,Gyr
\citep[e.g.,][]{bo78,bo84,dressler94,dressler97,treu,kodama}. 
Clusters at \mbox{$z\!\ga0.4$}
are dominated, particularly at faint magnitudes, by blue spiral
galaxies, predominantly irregular or Sc--Sd spirals. Some of these
show signs of disturbed morphologies, and many present spectroscopic
evidence that they have undergone multiple star-formation events over
the last \mbox{1--2\,Gyr} \citep{dressler94}. Conversely, local clusters are completely dominated
by passive early-type galaxies: elliptical and lenticular (S0) galaxies at the
brighter end, and dwarf spheroids at fainter magnitudes. 

In the standard hierarchical cosmological model, \mbox{$z\!\sim\!0.4$} represents
the peak infall rate of field galaxies onto the cluster \citep{kauffmann95}, and it is the
transformation of these infalling field galaxies from star-forming
disk-dominated galaxies into passively-evolving
spheroids over a period of \mbox{4--5\,Gyr} through their encounter with the
cluster environment, that produces the observed changes in cluster
galaxy populations. Several physical mechanisms related to the cluster
environment have been proposed as producing the observed
transformations in galaxies, in which interactions with either other
cluster galaxies or the hot intra-cluster medium (ICM) affect both
their structural and star-formation properties. 

To distinguish between
these mechanisms it is necessary to examine both where the transformations
occur, and how the star-formation and structural properties of the
galaxies are
changed \citep[e.g.][]{treu}. For example, ram pressure from the passage of the galaxy
through the dense ICM can effectively remove the cold gas supply and
thus rapidly terminate new star-formation, either by stripping the
gas directly \citep*{a99}, or by inducing a starburst in which all of the gas is consumed
\citep{fujita}. 
The most dramatic ICM-galaxy interactions
should occur when two clusters merge, as shock fronts created in the
ICM may trigger starbursts in galaxies over large scales \citep*{roettiger}.
Importantly, in terms of their environmental effects, these mechanisms
all require a dense ICM, and so their evolutionary effects are limited
to the cores of clusters. 

In contrast, galaxy mergers, which can strongly
affect the morphological evolution of disks, cannot occur when the
encounter velocities are much greater than the internal velocity
dispersion of galaxies \citep{aarseth}, and so while frequent in
small groups, are rare in rich clusters \citep{ghigna}.
Alternatively galaxy harassment, whereby repeated
close \mbox{($<$50\,kpc)}, high-velocity \mbox{($>$1000\,km\,s$^{-1}$)} encounters
with massive galaxies and the cluster's tidal field cause impulsive
gravitational shocks that damage the fragile disk of late-type
spirals \citep{moore}, transforming them over a
period of several Gyr. Galaxy harassment is effective throughout a
cluster, including beyond the virial radius, but its effects should be
greater for those clusters with higher velocity dispersions.
 
Finally, when a galaxy falls into a more massive halo, the
diffuse gas in its halo is lost to the ICM, thus preventing further
cooling and replenishment of the cold gas supply, ``suffocating'' the
galaxy \citep[e.g.][]{blanton00,diaferio}. Star-formation in the
galaxy then declines slowly as the remaining cold gas is used up \citep*{larson}.  

The large datasets provided by the 2dFGRS and SDSS have allowed the
environmental effects on galaxy properties to be followed
statistically over the full range of environments, from the sparse
field to the dense cluster cores \citep[e.g.][]{lewis,gomez}, at least
for the brightest galaxies (\mbox{${\rm M}<{\rm M}^{*}\!+1$}). They
show that star-formation is most closely dependent on local density,
and is systematically suppressed above a critical value of density,
that is found 3--4 virial radii from clusters, but also in galaxy
groups as poor as \mbox{$\sigma\sim100$\,km\,s$^{-1}$}. This
suppression is observed to be independent of the richness of the
structure to which the galaxy is bound \citep{tanaka}, indicating that
mechanisms such as galaxy harassment or ram-pressure stripping are not
important for the evolution of bright galaxies. Instead the strongest
candidates for driving their transformation are galaxy suffocation and
low-velocity encounters, which are effective in both galaxy groups and
cluster infall regions. 
 
However, it is not clear if and how this scenario extends to fainter
magnitudes, as there has been observed a strong bimodality in the
properties of galaxies about a characteristic stellar mass
\mbox{$\sim3\times10^{10}{\rm M}_{\odot}$} (corresponding to \mbox{$\sim{\rm
 M}^{*}\!+1$}), with more massive galaxies
predominately passive red spheroids, and less massive galaxies tending
to be blue star-forming disks \citep{kauffmann03}. This bimodality
implies fundamental differences in the formation and evolution of
giant and dwarf galaxies, and it has been proposed \citep[e.g.][]{dekel,keres} that these are
driven by thermal processes in the gas inflowing from the halo onto
the galaxy, with the characteristic mass scale representing the
point at which shocks in the halo become stable, heating up the halo
gas, and preventing further cooling. Hence if the formation and
evolution of giant and dwarf galaxies are so fundamentally different,
then they are likely to be affected differently by mechanisms related
to the environment. For example, galaxy harassment should be most
efficient at transforming low-luminosity late-type
galaxies. \citet{tanaka} find differences in the environmental
dependences of faint \mbox{(${\rm M}^{*}\!+1<{\rm M}<{\rm M}^{*}\!+2$)} and
bright \mbox{(${\rm M}<{\rm M}^{*}\!+1$)} galaxy populations, and suggest that
faint galaxies are affected by mechanisms related to the structure in
which the galaxy is found. 

To understand the mechanisms underlying the transformation of faint
galaxies requires datasets reaching much fainter luminosities. Crucial
discriminators between the different transformation mechanisms are the
time- and distance-scales involved: while ram-pressure stripping
should rapidly terminate star-formation in a galaxy within \mbox{$\la\!
100$\,Myr} but requires the dense ICM of the cluster core; a galaxy
undergoing suffocation will have its star-formation rate slowly
decline over a period of several Gyr. Hence the nature of the
transition from regions where the majority of galaxies are
star-forming, and those dominated by passive galaxies, will depend
strongly on the dominant mechanism involved. 

 One
approach is to use galaxy colours, which can be readily obtained to
much fainter magnitudes than spectroscopic star-formation rates, and which
through the use of models can be
directly related to star-formation histories with minimal assumptions
\citep[e.g.][]{bruzual}.  
Recent large datasets have shown that the bimodality of galaxies is also 
manifested through their broadband photometry, in
particular a separation can be made on the basis of colour into red
and blue galaxy populations \citep{strateva,blanton03}, which correspond
roughly to the two broad types previously known from their
morphological and spectroscopical characteristics: passively-evolving
early-type and star-forming late-type galaxies. This bimodality has
been further quantified, resulting in colour-magnitude (C-M) relations and for both the red
and blue galaxy populations \citep{baldry}, and its evolution observed
to \mbox{$z\sim1$} \citep{bell}. \citet{balogh04} show that the bimodal galaxy
colour distribution is strongly dependent on environment, with the
fraction of galaxies in the red distribution at a fixed luminosity
increasing from 10--30\% in the lowest density environments, to
\mbox{$\sim$70\%} at the highest densities. 

The most dramatic effects of environment on galaxy evolution should
occur in superclusters, where the infall and encounter velocities of
galaxies are greatest \mbox{($>$1000\,km\,s$^{-1}$)}, groups and clusters
are still merging, and significant numbers of galaxies will be
encountering the dense ICM of the cluster environment for the first
time.

With this in mind we are undertaking the Shapley Optical Survey (SOS), an
optical photometric study of the core region of the Shapley
supercluster \citep{shapley}, one the most massive structure in the local universe, containing as many as 25 Abell
clusters. 

In this paper we examine the effect of the supercluster environment on
the star-formation histories of galaxies as measured through their
galaxy colours. We present the SOS in Section~\ref{data}, and
then describe how we quantify the local environment and statistically
subtract the field galaxy population in Sections~\ref{den}
and~\ref{sub}. We present our analysis of the C-M relation in
Section~\ref{cmrelation}, which then allows us to separate the red and
blue galaxy populations whose spatial distributions are presented in
Section~\ref{redandblue}. We examine the environmental dependencies on
galaxy colours in Section~\ref{colours}, and discuss our findings in
Section~\ref{discussion}, before presenting our conclusions in Section~\ref{conclusions}. Throughout the paper we adopt a cosmology
with \mbox{$\Omega_M=0.3$}, \mbox{$\Omega_\Lambda=0.7$} and
\mbox{H$_{0}$=70\,km\,s$^{-1}$Mpc$^{-1}$}. According to this cosmology 1
arcmin corresponds to 60\,kpc at \mbox{$z=0.048$}.
 
\section{The Shapley Optical Survey}
\label{data}

The SOS comprises wide-field optical
imaging of a \mbox{2.0\,deg$^2$} region covering the whole of the
Shapley supercluster core (hereafter SSC), which comprises three Abell
clusters A3556, A3558 and A3562 and two poor clusters
SC\,1327-312 and SC\,1329-314. The resultant galaxy catalogues are complete to
\mbox{$R=22.0$} and \mbox{$B=22.5$}, allowing us for the first time to study the
effect of the supercluster environment on the photometric properties
of galaxies, particularly the dwarf galaxy population where we reach
\mbox{${\rm M}^{*}\!+7$}.

The area is covered by extensive redshift surveys
\citep*[e.g.][]{b00,quintana,drinkwater} which indicate that these clusters
form a complex clumpy and highly-elongated structure
\mbox{$\sim9\,h_{70}^{-1}$\,Mpc} across, which is in the final stages
of collapse, with infall velocities reaching
\mbox{2\,000\,km\,s$^{-1}$} \citep{reisenegger}. There exists a wealth
of multi-wavelength data from radio \citep[e.g.][]{miller} to X-ray
\citep[e.g.][]{finoguenov}, which in conjunction with the redshift data,
 describe a dramatic scenario with ongoing cluster-cluster mergers
 triggering star-formation in galaxies over $\sim$\,Mpc scales.
It is clear that amid this maelstrom of galaxies, groups and clusters, 
both galaxy harassment and ram pressure stripping should be at their
most effective in transforming the infalling field galaxies. 

We aim to use the SOS dataset to study the effect of environment on
the luminosity distribution, colours and structural properties of
galaxies, and through comparison with numerical simulations and
theoretical predictions, draw insights regarding which physical
processes are most important for the transformation and evolution of
galaxies in these environments.

In \citet[][hereafter Paper I]{paper1} we introduce the survey,
and describe the observations, calibrations, and derivation of the galaxy
catalogues. The galaxy luminosity functions are also presented, and
significant environmental effects are observed, in the form of a dip
at \mbox{$\sim{\rm M}^{*}\!+2$} which becomes deeper, and a faint-end
slope which becomes steeper, with decreasing density. We explain these
results in terms of the galaxy harassment scenario, in which the late-type
spirals that represent the dominant population at \mbox{$\sim{\rm
    M}^{*}\!+2$} are transformed by galaxy harassment into passively-evolving dwarf spheroids, and in the process become
\mbox{$\sim$1--2} magnitudes fainter due to mass loss and an ageing stellar population without new star-formation. 

In this paper we examine the effect of the supercluster environment on
the star-formation histories of galaxies as measured through their
galaxy colours, while in future papers we expect to consider the environmental
impact on galaxy structural parameters and compare our
observational results with semi-analytical models of galaxy evolution
in n-body simulations of a supercluster region.

\begin{figure*}
{\resizebox{17.7cm}{!}{\includegraphics{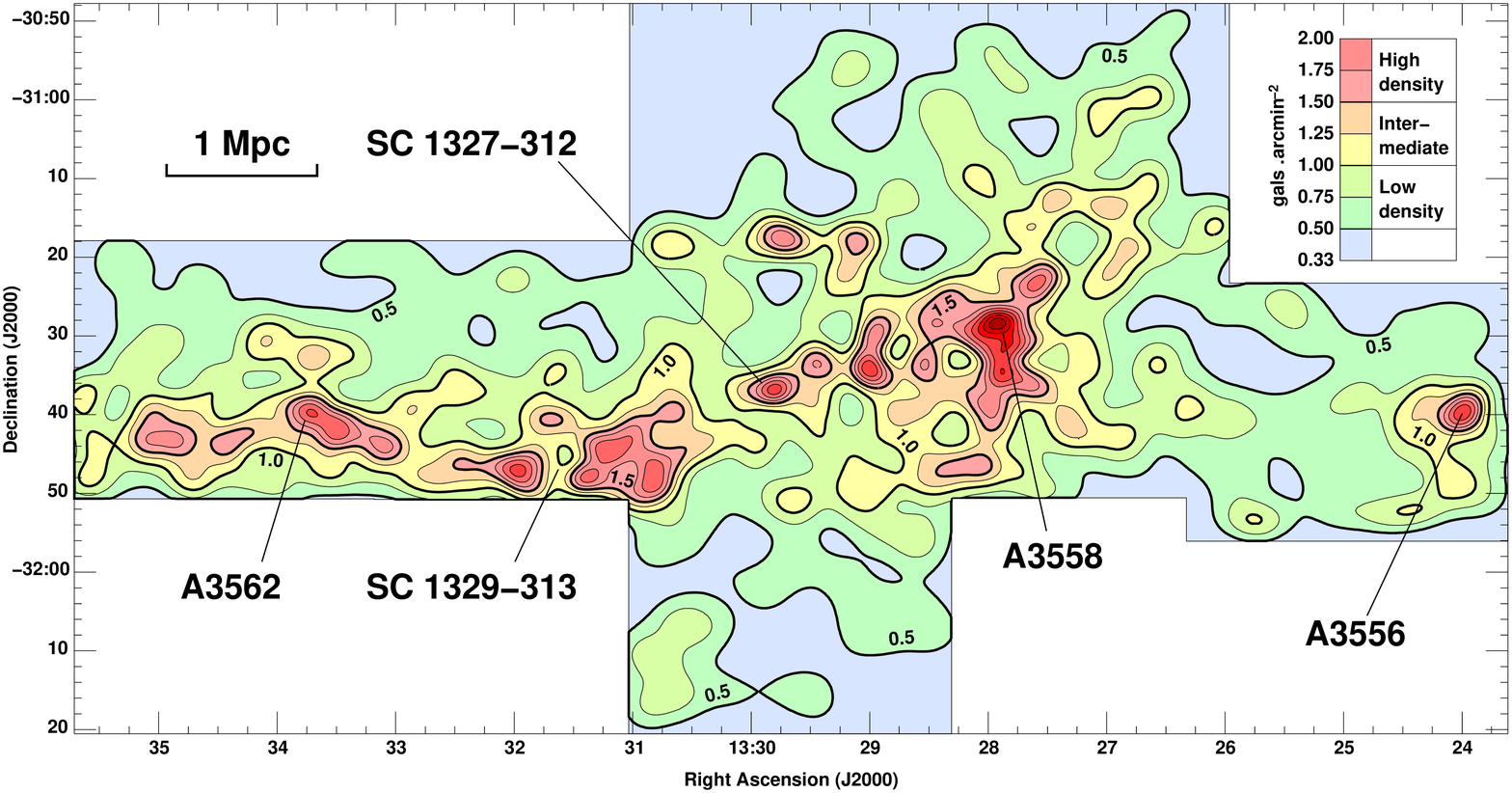}}}
\caption{The surface density of \mbox{$R<21.0$} galaxies in the region of the
  Shapley Supercluster core complex. Isodensity contours are shown at
  intervals of \mbox{0.25\,gals\,arcmin$^{-2}$}, with the thick contours 
corresponding to 0.5, 1.0 and \mbox{1.5\,gals\,arcmin$^{-2}$}, the
  densities used to separate the three cluster
  environments. High-density regions \mbox{$(\Sigma>1.5)$} are indicated by red
  colours, while intermediate- \mbox{$(1.0<\Sigma<1.5)$} and
  low-density \mbox{$(0.5<\Sigma<1.0)$} regions are indicated by
  yellow/orange and green colours respectively. The centres of X-ray
  emission for each of the clusters are indicated.} 
\label{densitymap}
\end{figure*}

The data were obtained from the ESO science archive
(68.A-0084, P.I. Slezak), comprising wide-field $B$- and $R$-band
imaging covering a \mbox{2.0\,deg$^2$} region towards the clusters A3562,
A3558 and A3556 which form the core of the Shapley supercluster at
\mbox{$z\sim0.05$}. Full details of the observations, data reduction,
and the production of the galaxy catalogues are described in Paper I,
and only a summary is presented below.

The observations were made from March 2002 to April 2003
using the WFI camera, an instrument made up of eight \mbox{$2046\times2048$}
CCDs giving a field of view of \mbox{$34'\times33'$}, and which is
set at the Cassegrain focus of the 2.2-m MPG/ESO telescope at La Silla.  
The survey is made up of eight contiguous fields, each with total exposure
times of 1\,500\,s in $B$ and 1\,200\,s in $R$, and typical FWHMs of
0.7--1.0$''$. The data are reduced using the ALAMBIC pipeline \citep[version 1.0,][]{vandame}, and calibrated to the Johnson-Kron-Cousins photometric
system using observations of \citet{landolt} standard stars. The
sources are then extracted and classified using SExtractor \citep{bertin}, resulting in galaxy catalogues which are
both complete and reliable (i.e. free of stars) to \mbox{$R=22.0$} and
\mbox{$B=22.5$}. For this analysis we consider only galaxies to \mbox{$R=21.0$} where
uncertainties in the photometry are less than \mbox{0.1\,mag} in both $R$ and
$B-R$ for galaxies belonging to the supercluster.

\section{Quantifying the Galaxy Environment}
\label{den}

To study the effect of the cluster environment on galaxies in the
SSC, the local density of galaxies, $\Sigma$, is
determined across the SOS mosaic. This is achieved using an adaptive
kernel estimator \citep{pisani93,pisani96}, in which each galaxy $i$ is
represented by a Gaussian kernel,
\mbox{$K(r_i)\propto\exp(-r^{2}/2\sigma_{i}^{2})$}, whose width, $\sigma_{i}$ is
proportional to \mbox{$\Sigma_{i}^{-1/2}$} thus matching the resolution
locally to the density of information available. For this study, we
consider the
surface number density of \mbox{$R<21.0$} \mbox{($<{\rm M}^{*}\!+6$)} galaxies, with an
additional colour cut applied to reject those galaxies more than 0.2\,mag redder in \mbox{$B-R$} than the observed cluster red sequence to minimize
background contamination. As there are no known structures in the
foreground of the SSC (90\% of \mbox{$R<16$}
galaxies have redshifts confirming that they belong to the
supercluster), any substructure identified in the density map is
likely to be real and belonging to the supercluster.
The local density is initially determined using a fixed Gaussian
kernel of width 2 arcmin, and then iteratively recalculated using
adaptive kernels. The resultant surface density map of the SSC
is shown in Fig~\ref{densitymap}, with the three clusters and two groups
indicated. Isodensity contours are shown at intervals of \mbox{0.25\,gals\,arcmin$^{-2}$}, with the thick contours corresponding to 0.5, 1.0 and
\mbox{1.5\,gals\,arcmin$^{-2}$}, the densities used to separate the three
cluster environments described below. 

The expected density of field galaxies is
 estimated through analysis of thirteen \mbox{$35'\times35'$} fields of deep
 $BVR$ imaging taken from the Deep Lens Survey \citep[DLS;][]{wittman}. These data covering \mbox{4.3\,deg$^2$} in total, were taken
 using the Mosaic-II cameras on the NOAO KPNO and CTIO 4-m telescopes,
 and have 5$\sigma$ depths of \mbox{$B,V,R\sim27$} and typical $R$-band
 FWHMs of 0.9$''$, allowing accurate photometric measurements and
 star-galaxy classifications to be made to at least the depths of our
 survey. Through applying the same colour-magnitude cuts, we estimate
 the density of field galaxies to be \mbox{$0.335\pm0.019$\,gals\,arcmin$^{-2}$}, 
and hence the thick contours
correspond to overdensity levels of \mbox{$\sim50$}, 200 and \mbox{400\,gals\,$h_{70}^{2}$\,Mpc$^{-2}$} respectively. The entire region
covered by the SOS can be seen to be overdense with respect to field galaxy
counts.

For the following analyses on the
effect of the cluster environment on its constituent galaxy population
we define three regions selected according the local surface number
density. Firstly we consider a high-density region with \mbox{$\Sigma>1.5$}\,gals\,arcmin$^{-2}$ which correspond to the cores of the clusters. Next
we consider intermediate- (\mbox{$1.0<\Sigma<1.5$}) and
low-density (\mbox{$0.5<\Sigma<1.0$}) regions which probe
the filament connecting the clusters A3562 and A3558, as well as the
wider envelope containing the whole supercluster core complex. 

\section{Statistical Field Galaxy Subtraction}
\label{sub}

To accurately measure the global photometric properties of the galaxy
population in the SSC requires the
foreground / background contamination to be estimated efficiently and
corrected for. There exists already a wealth of spectroscopic data
in the region, comprising 607 published galaxy redshifts \citep{b98} corresponding to
90\% of \mbox{$R<16$} galaxies, dropping to 50\% for
\mbox{$R<17.7$}. For those galaxies without redshifts, particularly at
fainter magnitudes, we estimate the probability that they are
supercluster members. This probability is dependent on the spatial
position, the $R$-band magnitude and $B-R$ colour. 
The dependence on its spatial position, through its local number
density, is clear as galaxies towards one of the
density peaks will be more likely to be members than those in regions where
the surface density is closer to that expected for the field. The
dependence on the galaxies colour and magnitude is complex, with
galaxies located on the cluster red sequence most likely to be
members, while those galaxies much redder than the sequence most
likely to belong to the background population. It is important that
all three parameters are considered simultaneously, as the
relationship between the broad-band properties of galaxies and their
environment has been shown to be complex \citep{blanton03}. 

We account for this by considering separately the three cluster
environments described previously when estimating the probability
that galaxies are supercluster members through their $R$-band
magnitude and $B-R$ colour using the methods described in \citet{kodama01}.
For each of these three cluster environments and the Deep Lens Survey field
comparison, two-dimensional histograms
are built with bins of width 0.4\,mag in $R$ and 0.2\,mag in $B-R$ to
properly map the galaxy C-M distribution. The
histogram of field galaxies is normalised to match the area within
each cluster region, and for each galaxy, the probability that galaxy $i$
belongs to the supercluster is then defined as:
\begin{equation}
P_{SC}^{i}=\frac{\Sigma_{SC}(R,B-R)_{i}-\Sigma_{field}(R,B-R)_{i}}{\Sigma_{SC}(R,B-R)_{i}}, 
\label{psc}
\end{equation}
where \mbox{$\Sigma_{SC}(R,B-R)_{i}$} and \mbox{$\Sigma_{field}(R,B-R)_{i}$} are the number
densities of galaxies in the supercluster and field environments
belonging to the corresponding bin in $R$ and $B-R$. In low-density
environments it is possible that
\mbox{$\Sigma_{SC}(R,B-R)_{i}<\Sigma_{field}(R,B-R)_{i}$} for a particular bin, and
  so in these cases the excess number of field galaxies are
  redistributed to neighbouring bins with equal weight until
  \mbox{$\Sigma_{SC}(R,B-R)_{i}\geq\Sigma_{field}(R,B-R)_{i}$} for all bins. 
For those galaxies with redshifts \mbox{$P_{SC}^{i}$} is set to 1 for
\mbox{$0.035<z<0.056$} (the redshift range of known supercluster
members in the SOS field) or 0 otherwise. Hence for each \mbox{$R<21$} galaxy in the
region covered by the SOS, we estimate the likelihood of its
belonging to the Shapley supercluster based upon its $R$-band
magnitude, $B-R$ colour and local density.

\begin{figure}
\centerline{{\resizebox{9cm}{!}{\includegraphics{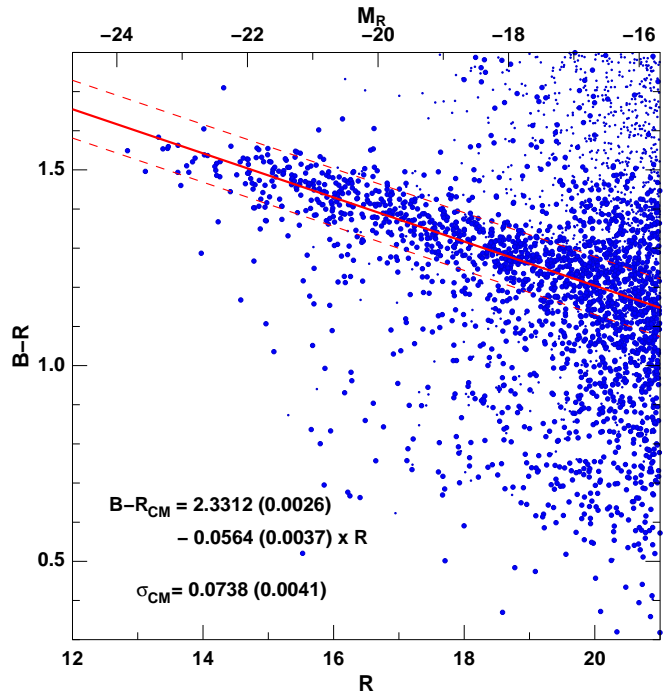}}}}
\caption{The $B-R/R$ colour-magnitude diagram of galaxies in the
  regions corresponding to intermediate- and high-densities
  (\mbox{$\Sigma>1.0$}\,gal\,arcmin$^{-2}$). The larger points represent a
  Monte Carlo realisation of the supercluster galaxy population, while
  the remaining (and hence field) galaxies are  shown by the small
  points. The best-fitting C-M relation is indicated by the solid red
  line, with the parallel dashed lines indicating the $1\sigma$
  dispersion levels.} 
\label{CM}
\end{figure}

\section{The Colour-Magnitude relation}
\label{cmrelation}

To estimate the location of the colour-magnitude relation for galaxies
in the supercluster, we consider those galaxies with local densities
\mbox{$\Sigma>1.0$}, i.e. both intermediate- and high-density
environments, in order to avoid {\em a priori} field contamination
effects when measuring the C-M slope. Figure~\ref{CM} shows the resulting C-M diagram for a Monte
Carlo realisation of the galaxy population in these regions, in which
those galaxies predicted to belong to the supercluster indicated by
the larger symbols, and those belonging to the field indicated by the
small points. A clear red sequence is apparent for \mbox{$13<R<20$}. However
at \mbox{$R<15$} it becomes notably flatter, while at \mbox{$R>19$} it becomes
increasingly difficult to separate it from the remainder of the
cluster population. This flattening at the brightest magnitudes has
been noted previously for this region \citep{metcalfe}, and has
been observed for a large sample of galaxies from the Sloan Digital
Sky Survey \citep{baldry}, indicating this is a universal
phenomenon. 
The slope in the C-M relation has
been considered due to a progressive increase in metallicity with mass
due to the greater ability of more massive galaxies to retain metals
dispersed by supernova during their formation, and so the observed
flattening of the C-M slope probably represents the same flattening of the
mass-metallicity seen for galaxies with
\mbox{${\rm M}_{*}\!\ga\!10^{10.5}\,{\rm M}_{\odot}$} \citep{tremonti}, which in
turn may reflect the observation that the brightest galaxies are built
up through ``dry mergers'' of \mbox{$\sim\!{\rm M}^{*}$}
galaxies \citep[e.g.][]{faber}, and hence would have the same metallicities and colours as their
progenitors.

We perform a fit to the C-M relation over the range \mbox{$14.5<R<19.0$},
where the red sequence relation appears linear and can be separated
from the rest of the cluster population. Although the red sequence
appears visually quite distinct, obtaining a robust fit to it is not
trivial, particularly given the presence of outliers which are heavily
skewed bluewards of the relation, and so simple least-squares methods
are unsuitable. We apply the same method as \citet{lopez}, which produced
 the most robust results by iteratively
applying the biweight algorithm \citep*{beers} with 3$\sigma$
clipping to counteract the heavily skewed distribution. A range of
slope values were searched to determine the best-fitting value that
minimizes the biweight scale of the deviations from the median value.

The uncertainties in the slope and intercept are derived using the
non-parametric bootstrap method. The number of cluster galaxies, $n$, is
estimated as \mbox{$n=\sum_{i=1}^{N} P_{SC}^{i}$} for the $N$ galaxies with \mbox{$14.5<R<19$}, and each
bootstrap performed by sampling $n$ galaxies with replacement according to
the probabilities $P_{SC}^{i}$ calculated using Eq.~\ref{psc}. \citet{babu} have analytically
estimated that \mbox{$\sim n\log^{2}n$} bootstrap resamplings give a good
approximation to the underlying density population. The resultant
best-fitting C-M relation is found to be:
\begin{equation}
(B-R)_{CM} = 2.3312\pm0.0026 - 0.0564\pm0.0037 \times R,
\label{cmequation}
\end{equation}
where the quoted uncertainty in the intercept assumes a fixed value of
slope. The obtained dispersion about the relation as given by the
biweight scale of the deviations is \mbox{$0.0738\pm0.0041$}\,mag.

\begin{figure}
\centerline{{\resizebox{9cm}{!}{\includegraphics{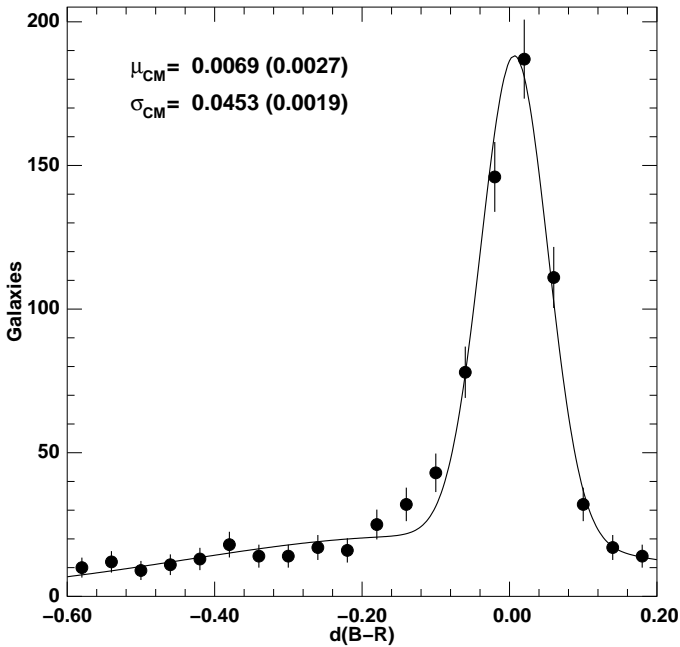}}}}
\caption{The distribution of $B-R$ galaxy colours offsets from the
  C-M relation predicted by the biweight algorithm for \mbox{$R<19$} galaxies
  with local densities greater than 1.0\,gal\,arcmin$^{-2}$.}
\label{CMdist}
\end{figure}

As a robust test of the results obtained using the biweight algorithm, we plot a histogram of the distribution of
\mbox{$\Delta (B-R)_{CM}$}, the $B-R$ galaxy colour
offset from the C-M relation, and model the distribution as a sum of
two Gaussian functions, one to represent the red sequence, and
another to represent the ``blue cloud'' population. Fig~\ref{CMdist}
shows that a bimodal Gaussian distribution describes well the
distribution of \mbox{$\Delta (B-R)_{CM}$} \mbox{($\chi^{2}/\nu\approx1$)}, but with a mean offset of
\mbox{$0.0069\pm0.0027$\,mag} with respect to the biweight median, and a
dispersion of \mbox{$0.0453\pm0.0019$\,mag} that is significantly smaller
than predicted by the biweight scale estimator, and comparable to
the \mbox{$0.054$\,mag} dispersion observed for the Coma cluster in the same
passbands \citep{lopez}. Hence despite iteratively clipping galaxies
\mbox{$>3\sigma$} from the median value of the relation, both the intercept
and dispersion of the relation obtained using the biweight algorithm have been affected by outliers and the heavily skewed distribution. 

We do not expect that photometric uncertainties are a major
contributor to the observed dispersion as: firstly the mean $B-R$ uncertainty
for galaxies in the red sequence is found to be 0.012\,mag, only
reaching 0.025\,mag by \mbox{$R=19.0$}; and secondly the observed dispersion
does not vary significantly as a function of magnitude over \mbox{$14.5<R<19$}.

It is remarkable that such a small dispersion is obtained, given the
large number of galaxies involved \mbox{($\sim600$)} over two orders of
magnitude in luminosity, and spread across three rich clusters and
\mbox{$\sim 9\,h_{70}\,$Mpc}. The wide spread of redshifts within the
supercluster complex \mbox{($0.035<z<0.056$)} could introduce a significant
component to the observed dispersion through the spread of
k-corrections required for the red sequence galaxies, as the predicted
change in the $B-R$ k-correction of an elliptical galaxy from $0.035$
to $0.056$ is \mbox{$\sim0.08$\,mag} \citep{poggianti}, as is the observed
increase in $B-R$ colour of red sequence galaxies \citep{lopez}.

To measure the possible effect of k-corrections on the obtained colour
dispersion of red sequence galaxies, we consider those red sequence
galaxies with known redshifts. Although the redshift data is less than
50\% complete beyond \mbox{$R=17.7$}, the $B-R$ colour dispersion of the red
sequence galaxies with redshifts is consistent with that for the whole
\mbox{$R<19$} sample. After applying suitable k-corrections to each of the galaxies
with known redshifts, and the colour dispersion about the red sequence
is recalculated, found to be still consistent with that before the
k-correction has been applied. Hence the observed internal dispersion
of galaxy colours around the red sequence is not primarily the effect
of different k-corrections over the redshift range of the supercluster.
This
can be explained by considering that the redshift distribution has a
standard deviation of 0.0036 \mbox{($\sim1\,100\,$km\,s$^{-1}$)},
corresponding to an rms $\sigma$ of $\sim0.014$\,mag in $B-R$,
significantly smaller than the observed $\sigma_{CM}$. 
  
\begin{figure*}
{\resizebox{17.7cm}{!}{\includegraphics{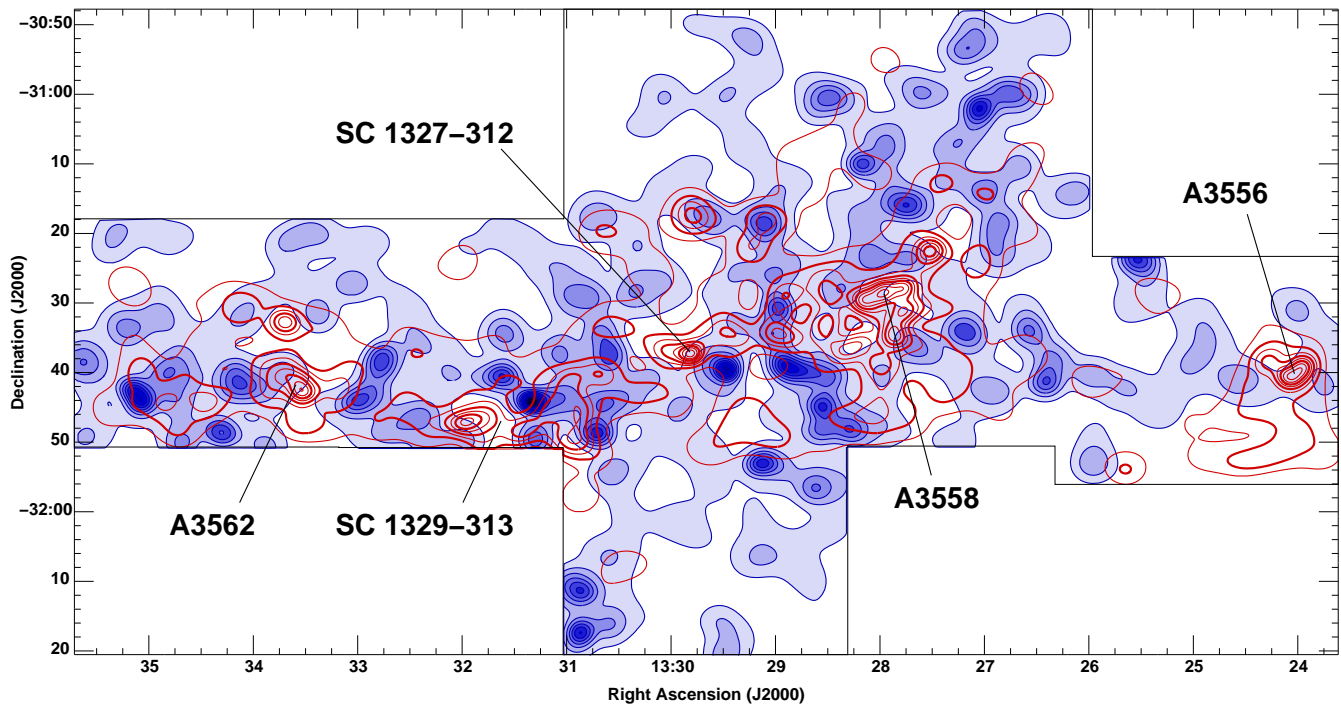}}}
\caption{The surface density of \mbox{$R<21$} galaxies \mbox{$>3\sigma$} bluer than the
  cluster red sequence (blue filled contours) and those
  within \mbox{$3\sigma$} of the cluster red sequence (red contours) over the
  supercluster core complex. The isodensity contours for the blue
  galaxies are shown at intervals of 0.125\,gals\,arcmin$^{-2}$,
  while the red contours are at intervals of 0.25\,gals\,arcmin$^{-2}$. The centres of X-ray emission for each of the
  clusters are indicated.}
\label{bluegals}
\end{figure*}

\begin{table}
\begin{tabular}{cccc}
$R<21$ gals & & & \underline{$\chi^{2}$} \\
arcmin$^{-2}$ &  $\mu_{CM}$ & $\sigma_{CM}$ & $\nu$ \\ \hline
$\Sigma>1.0$ & $+0.0069\pm0.0027$ & $0.0453\pm0.0019$ & 0.99 \\ \hline
$\Sigma>2.0$ & $+0.0191\pm0.0064$ & $0.0460\pm0.0044$ & 0.24 \\
$\Sigma>1.5$ & $+0.0142\pm0.0043$ & $0.0486\pm0.0031$ & 0.47 \\
$1.0<\Sigma<1.5$ & $-0.0004\pm0.0035$ & $0.0442\pm0.0024$ & 1.40 \\
$0.5<\Sigma<1.0$ & $-0.0012\pm0.0042$ & $0.0540\pm0.0031$ & 1.04 \\ \hline
\end{tabular}
\caption{Gaussian fits to the $B-R$ colour distribution around the C-M
  relation for \mbox{$R<19$} galaxies in the different density environments}
\label{gauss}
\end{table}

To examine the effect of environment on the colour of the C-M relation
we fit
bimodal Gaussian distributions to the \mbox{$\Delta (B-R)_{CM}$} distribution
in the three environments as parametrised by the local \mbox{$R<21$} galaxy
density. The slope and zero-point are maintained fixed at the values
indicated in Eq.~\ref{cmequation}, in order to measure the shifts in
the overall $B-R$ colour of the red sequence with
environment. We do not make any attempt to determine the slope and
intercept in the differing environments using the biweight algorithm,
as the effect of the heavily skewed distribution on the determination
of these values will be dependent on environment, due to the
increasing blue galaxy fraction from high- to low-density
regions. Hence the resulting fits would be biased by the changing blue
galaxy population, and mask any inherent changes in the red sequence
population with environment.

The results of the Gaussian fits to the red sequence population are
shown in Table~\ref{gauss}. In each case the galaxy colour
distribution is well described by a double Gaussian function.
The zero-points of the red sequence are
consistent for the low- and intermediate-density regions, but the
red sequence is found to be \mbox{$0.0147$\,mag} redder in the high-density
region, a result significant at the \mbox{2.6$\sigma$} level. The redward
shift of the red sequence continues if the density
threshold is increased further, reaching \mbox{$0.0195$\,mag} for
\mbox{$\Sigma>2.0$}. 
The $B-R$ colour dispersion of the red sequence remains
constant at \mbox{$\sim$0.045}, for each of the environments, suggesting that
the observed reddening indicates an overall {\em shift} in the red
galaxy population, rather than a {\em broadening} of the distribution
due to increased numbers of bluer galaxies.   

\section{The Spatial Distribution of Red and Blue Galaxies}
\label{redandblue}

\begin{figure*}
{\resizebox{17.7cm}{!}{\includegraphics{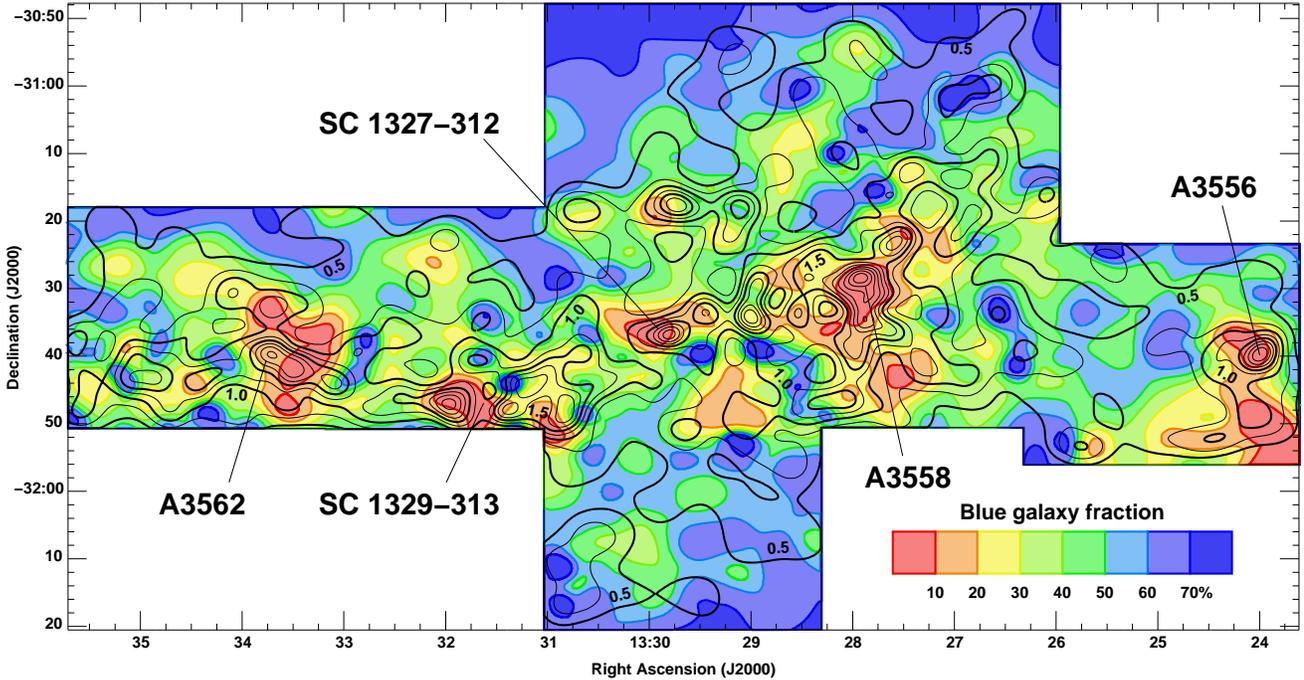}}}
\caption{The fraction of $R<21$ galaxies $>3\sigma$ bluer than the
  cluster red sequence as a function of spatial position (see text).}
\label{bluegalfrac}
\end{figure*}

To examine where star-formation has been triggered or inhibited by the
supercluster environment, we split the galaxy population into two
according to colour. We identify red galaxies as those within
$3\sigma$ (after including photometric uncertainties) of the cluster red sequence, and blue
galaxies as those \mbox{$>3\sigma$} bluer than the red sequence. Hence these
blue galaxies are all those {\em not} belonging to the red sequence,
due to being either currently having some star-formation, or having had some
in the previous \mbox{$\sim$2\,Gyr}. 

Figure~\ref{bluegals} shows the spatial distribution of both red and blue galaxies
over the supercluster complex, after correcting for the expected
background contribution of 0.147 and \mbox{0.071\,gals\,arcmin$^{-2}$} for the
red and blue galaxy subsets respectively, as estimated from the 13
Deep Lens Survey fields. The blue isodensity contours for the blue galaxies
are shown at intervals of \mbox{0.125\,gals\,arcmin$^{-2}$}, which are
filled with increasingly intense blue colours with density. Overlaid
are the red isodensity contours at intervals of \mbox{0.25\,gals\,arcmin$^{-2}$} to represent the local density of the red galaxy population.
Note that all the contours indicate overdensities in comparison to the
field, and that the spacing of the contours for the red galaxy
population is twice that of the those for their blue counterparts.
The centres of the X-ray emission from each of the clusters in the
region are indicated.

The spatial distribution of the red and blue galaxy populations differ
significantly. Whereas the blue galaxies appear reasonably evenly
spread across the whole field, the red galaxies are highly
concentrated towards the cluster centres and the filamentary structure
connecting A3562 and A3558.
In contrast, none of the three Abell or two poor clusters
show particular overdensities in the blue galaxy distribution towards their
cores, in fact they appear underdense in comparison to their immediate
surroundings, as noted by \citet{metcalfe} in the case of A3558.
 However, several localised
overdensities of blue galaxies are apparent along the filamentary
structure connecting A3562 and A3558, indicative of regions of
enhanced star-formation within the supercluster core complex. What is
particularly interesting is their distribution within the filamentary
structure with respect to the clusters. The most overdense regions are
located {\em between} each of the clusters A3562, A3558 and the two poor
clusters. The two most notable overdensities of blue galaxies appear
to be the western flank of the poor cluster SC\,1329-313, and a linear
structure bisecting the clusters A3558 and SC\,1327-312. Both these
overdensities contain numerous blue galaxies which are spectroscopically
confirmed as belonging to the supercluster. The region where these
overdensities of blue galaxies are located is where the two clusters
A3562 and A3558 are both experiencing merging events, which may suggest that
this merging has recently triggered star-formation in these blue
galaxies, possibly through shock fronts produced when the ICMs of
merging clusters collide. Alternatively these overdensities could
reflect infall regions along the filament connecting the clusters
A3558 and A3562. 

Further insights into the effect of the supercluster on star-formation
in its member galaxies are possible through the complementary
information provided by the {\em fraction} of supercluster
galaxies that are classed as blue. Figure~\ref{bluegalfrac} shows the
fraction of \mbox{$R<21$} galaxies \mbox{$>3\sigma$} bluer than the cluster red
sequence as a function of spatial position, as obtained by dividing
the local surface density of blue galaxies by that for all \mbox{$R<21$}
galaxies after correcting both for background contamination. The blue
galaxy fraction is represented by the coloured contours, with red
indicating low blue galaxy fractions \mbox{($<10$\%)} and blue indicating
high blue galaxy fractions \mbox{($>50\%$)}. For comparison the isodensity
contours of the overall \mbox{$R<21$} galaxy density of Fig.~\ref{densitymap}
are overlaid.  

Whereas the local blue galaxy density shows little correlation with that of
the overall local galaxy density (see Fig.~\ref{bluegals}), the blue galaxy {\em fraction}
appears strongly anti-correlated with the overall local galaxy
density. Each of the three Abell clusters, and the two poor clusters,
are marked by regions of low \mbox{($<10\%$)} blue galaxy fractions, while
the filamentary structure connecting the clusters A3562 and A3558 has
typically \mbox{$\sim$20--50\%} blue galaxies, which rises still further for
the regions furthest from the clusters. It is however this avoidance of the
cluster cores by the blue galaxies that is the most notable feature of
Fig.~\ref{bluegalfrac}, with \mbox{$>90\%$} of galaxies in these regions
located within the narrow locus of the cluster red sequence to \mbox{$R=21$}
or \mbox{$M^{*}+6$}. 
It is clear that in the central regions \mbox{($r<300$\,kpc)} of
all the clusters in this region star-formation has been severely
truncated in {\em almost all if not all} galaxies. 

\begin{figure}
\centerline{{\resizebox{8.0cm}{!}{\includegraphics{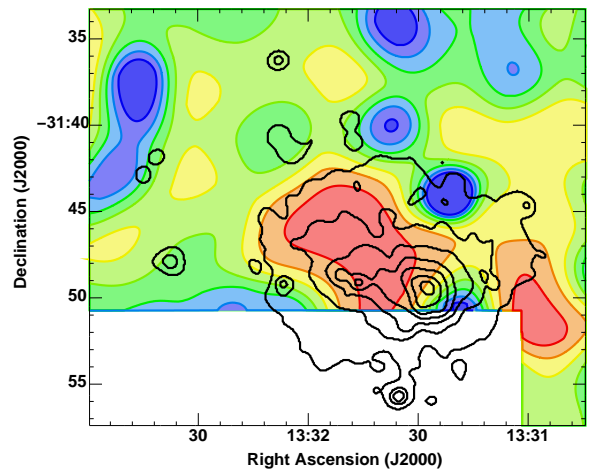}}}}
\caption{The relation between the fraction of blue galaxies (coloured
  filled contours as in Fig.~\ref{bluegalfrac}) and X-ray
  emission (black contours) in the vicinity of the cluster SC\,1329-313.}
\label{xmm}
\end{figure}

A crucial insight into what physical process is resulting in this
termination of star-formation can be obtained through comparison with
complementary XMM X-ray imaging of the region around A3562 and SC\,1329-313 as reported in \citet{finoguenov}. A comparison of
their Fig.\,1 shows that
while the centre of the X-ray emission from A3562 is coincident with
that obtained from our galaxy surface density, there is an extension
to the north-east, where we also find the blue galaxy fraction to be
lowest as shown in Fig.~\ref{bluegalfrac}. 

Figure~\ref{xmm} shows how the blue galaxy fraction varies with spatial
position in the vicinity of SC\,1329-313, and whereas in
Fig.~\ref{bluegalfrac} the overlaid black isodensity contours show the
local galaxy density distribution, here they represent the X-ray
emission from the ICM of SC\,1329-313, taken from the XMM science archive
(OID 10526, P.I. Aschenbach).

The centre of X-ray emission (as indicated by the arrow in Fig.~\ref{densitymap})
does not coincide well with either of the two peaks in the local
galaxy density distribution, but appears to lie in between them. This
displacement between the ICM and the galaxy distribution may
be a dynamical effect from the merger, the result of the collisional
gas lagging behind the collisionless DM and galaxies.
A comparison of Figs.~\ref{bluegals} and~\ref{xmm}
 suggests that while the region where the blue galaxy fraction
is lowest does not coincide with either of the two peaks in the local
galaxy distribution, it covers approximately the same region as the
observed X-ray emission from the cluster.
For both the clusters A3562 and SC\,1329-313, regions where the
blue galaxy fractions are lowest \mbox{($<20$\%)} appear more strongly
coincident with the centres and
extensions of the X-ray emission, than the corresponding centres and
extensions of the clusters as measured by the overall galaxy number
density. The regions of low blue galaxy fractions appear coincident
with the peaks in the galaxy surface density for the remaining
clusters A3558, A3556 and SC\,1327-312, and so it is not possible to
separate the effects of galaxy density with those of the ICM for these
regions.

\section{The Relations Between Colour, Luminosity and Environment}
\label{colours}

\begin{figure}
\centerline{{\resizebox{8.0cm}{!}{\includegraphics{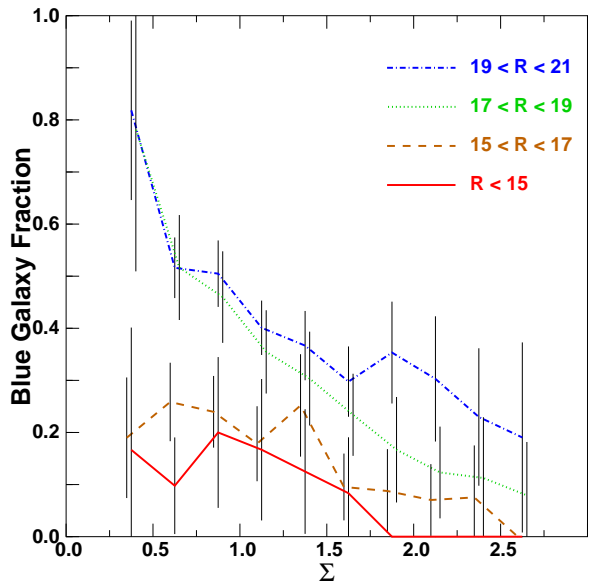}}}}
\caption{The blue galaxy fraction as a function of local
  density. Four lines are shown corresponding to four different \mbox{$R$-band} magnitude ranges as indicated.}
\label{blue_den}
\end{figure}

As galaxy colours can be directly related to star-formation
histories, by examining the environmental dependences of galaxy
colours it is possible to gain insight into where and how
star-formation is affected by the supercluster environment. 
Figure~\ref{blue_den} shows the overall blue galaxy fraction as a
function of local density for four different $R$-band magnitude
ranges, quantifying the global anti-correlation with local
density. The highest-density regions \mbox{($\Sigma\ga1.5$)} are dominated by
red sequence galaxies over the entire magnitude range
studied. There are two clear trends: the blue galaxy fraction
decreases with increasing density; and in all environments the blue
galaxy fraction increases monotonically with magnitude.  
The strength of this second trend appears
greatest at \mbox{$R\sim17$}: 
whereas at bright magnitudes \mbox{($R<17$} or
  \mbox{M$_{R}\!\la\!{\rm M}^{*}+2$}) the fraction
of blue galaxies only increases marginally with decreasing local
density, reaching only \mbox{$\sim20$\%} at the lowest densities studied; at
fainter magnitudes \mbox{($R>17$)} the blue galaxy fraction increases rapidly
to \mbox{$\sim80$\%} in the lowest density bin. 

\begin{figure}
\centerline{{\resizebox{8.5cm}{!}{\includegraphics{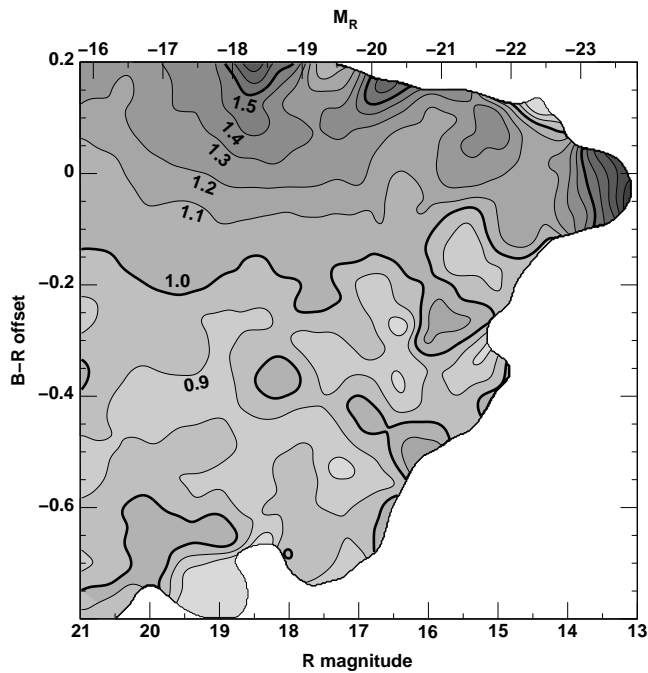}}}}
\caption{The mean local density as a function of the $B-R$ colour
  offset from the red sequence and $R$-band magnitude.}
\label{blanton}
\end{figure}

The most luminous members of the supercluster are predominantly red
sequence galaxies, and so would be expected to be strongly clustered
in the high-density regions.
 To examine whether luminosity segregation contributes to the
observed trends in galaxy colour with density, we plot in
Fig.~\ref{blanton} the average environment as a function of both galaxy
colour and magnitude.
The grey-scale and contours indicate mean local density as a function
of both the $B-R$ colour offset and $R$-band magnitude, smoothed on scales of
\mbox{0.05\,mag} in $B-R$ and \mbox{0.4\,mag} in $R$. The global trends are very
similar to those observed for galaxies in the SDSS \citep{blanton04,blanton05},
with the predominantly horizontal contours indicating that for all but the
brightest galaxies, where environment is strongly
luminosity-dependent, local density is a strong function of galaxy
colour, while remaining almost independent of luminosity.
Red galaxies are typically found in high-density regions, while blue
galaxies are on average located in low-density regions, the two
populations separated at a density \mbox{$\sim$1.0--1.1\,gals
arcmin$^{-2}$}, or \mbox{$\sim$200--250\,gals\,$h_{70}^{2}\,{\rm Mpc}^{-2}$}.
This separation reflects the trend shown in
Fig.~\ref{blue_den} for the blue galaxy
fraction to increase with decreasing local density.

In a comparable study of the A901/902 supercluster at
\mbox{$z=0.16$}, \citet{gray} use the 17-band COMBO-17 survey data to
precisely isolate supercluster members within a thin photometric
redshift slice \mbox{($0.15<z_{phot}<0.18$)}. They find a strong
segregation of faint quiescent and star-forming galaxies around a
local surface density of \mbox{$\sim200\,h_{70}^{2}\,{\rm Mpc}^{-2}$}
to a limit of \mbox{M$^{*}\!+6$}, in excellent agreement with our value. 
While \citet{gray} find a sharp transition in the galaxy properties at
this density, our observed trends in mean galaxy colour and blue
galaxy fractions with environment appear more gradual. This we put
down to the fact that whereas \citet{gray} use the
17-band photometry to accurately isolate the supercluster galaxies, we
perform a statistical background subtraction based on just the
$R$-band magnitude, $B-R$ colour and local density, and hence any
trends will be blurred out somewhat by interlopers.

We find that the brightest galaxies \mbox{($R<14$)} are the most
clustered, which could appear contrary to the results of
\citet{metcalfe} who found that \mbox{$b<17$} galaxies within 1.1\,Mpc of
A3558 were significantly less clustered than those with \mbox{$b>17$} both in
projection and redshift space. However their limit of \mbox{$b=17$}
corresponds to \mbox{$R\sim15.5$}, which is more than a magnitude fainter
than the point at which we find luminosity segregation to become important.

\citet{blanton04} also reported that, as well the environments becoming
denser with increasing luminosity for red galaxies at the brightest
magnitudes, the environments of low-luminosity red galaxies become denser with
increasing absolute magnitude down to \mbox{M$_{R}=-17.7$}, the limit of
their survey. A similar phenomenon is apparent in our data, and
although our survey extends two magnitudes fainter, we find that the
trend for low-luminosity red galaxies to be found in dense
environments is limited to those with \mbox{M$_{R}\sim-18$}. However it
should be remembered that the uncertainties from the statistical
background subtraction we have applied to measure the global
supercluster galaxy populations become increasingly large at faint magnitudes. 

\section{discussion}
\label{discussion}

We have examined the effect of the Shapley supercluster environment,
as measured in terms of the local surface density of \mbox{$R<21.0$}
galaxies, on the star-formation histories of galaxies through their
colours.

\subsection{The Colour-Magnitude Relation}

To examine the effect of environment on the {\em global} properties of
red sequence galaxies, we measure the {\em overall} distribution of
$B-R$ colours with respect to the observed C-M relation using a double
Gaussian fit to describe the global bimodal galaxy distribution. The
use of this approach to measure environmental trends is vital, as the
usual method of obtaining the best-fitting C-M relation and measuring
the effect of environment on these fits are biased by the strongly
skewed distribution and changing contribution of the blue galaxy
population. We show that the {\em location} of the C-M
relation is dependent on environment, with the relation
\mbox{$0.0147\pm0.0055$} redder in $B-R$ in the high-density regions
corresponding to the cluster core (\mbox{$r\siml0.3h_{70}$\,Mpc}), than the low- and
intermediate-density regions which represent the cluster periphery
(\mbox{$0.3\siml r\siml1.5\,h_{70}$\,Mpc}). This is a similar result to
that obtained for the rich cluster A209 at $z=0.21$ where the red
sequence was observed to be \mbox{$0.022\pm0.014$\,mag} redder in $B-R$ in
the cluster core than the periphery \citep{haines04}. By considering a
model red sequence galaxy as a 10\,Gyr old (at \mbox{$z=0$}),
\mbox{$\tau=0.01$\,Gyr}, solar-metallicity stellar population \citep{bruzual},
the observed reddening of the C-M relation in the high-density region
of A209 was interpreted as these galaxies being on average 500\,Myr
older or 20\% more metal-rich than their counterparts in lower density
environments. The observed reddening in the high-density regions of
the Shapley supercluster is also consistent with the red sequence
galaxies being \mbox{$\sim$500\,Myr} older in the high-density regions than
those in lower density environments, the resultant reddening of the
$B-R$ colour being 33\% lower than for A209 as the galaxies should be
on average 2\,Gyr younger in A209 than in the Shapley supercluster.

These results are consistent with analysis of the spectra of \mbox{22\,000}
luminous, red, bulge-dominated galaxies from the SDSS, which were
found to be marginally older and more metal-rich in high-density
regions than their low-density counterparts \citep{eisenstein},
although they were unable to quantify these differences due to their
models being unable to vary $\alpha$--Fe abundance ratios. By
considering the effects of age, metallicity and $\alpha$--Fe ratios
simultaneously on the spectra of 124 local early-type galaxies,
\citet{thomas} show that those in high-density environments are \mbox{$\sim$2\,Gyr}
older and slightly \mbox{(10--25\%)} more metal-rich than their counterparts
in field regions. We should expect much smaller differences in the
predicted average ages of red sequence galaxies as even our
low-density environments are typically within 1 cluster radii, while
gradients in the star-formation histories of cluster galaxy
populations have been shown to extend out to 3--4 virial radii before
reaching field values. 

These results are understandable in the framework of cosmological
structure formation models, in which galaxies form earliest in the
highest-density regions corresponding to the cluster cores, and have
their star-formation terminated earliest as the cores become filled
with shock-heated virialised gas which does not easily cool or
collapse, suppressing further formation of stars and galaxies
\citep{blanton00}. Age-gradients remain in clusters to this epoch, as
mixing of the galaxy population is incomplete, and the position of the
galaxies within the cluster are correlated with the epoch at which
they were accreted, and the point at which star-formation within them
is suppressed \citep{diaferio}.
 
\subsection{Galaxy Colours}

We identify two predominant trends relating galaxy colours,
luminosities and their local environment: the anti-correlation between
blue galaxy fraction and local density; and the monotonic increase in
the blue galaxy fraction with $R$-band magnitude in all environments
studied. These trends mirror those obtained by \citet{balogh04} relating
the red galaxy fraction (selected by $u-r$ colour), ${\rm M}_{r}$ and local density for galaxies taken from the SDSS
first data release, although they indicate that the strength of the
trend with environment is independent of luminosity, whereas we find
that the effect of environment on the blue galaxy fraction is much
greater at fainter magnitudes ($17<R<21$) than at brighter magnitudes
($R<17$) where red galaxies dominate in all environments. As all the
environments studied here lie within $\sim\!1\,{\rm R}_{vir}$ of one
of the clusters, this suggests that processes related to the
supercluster environment are much more important for the evolution of
faint galaxies ($\ga\!{\rm M}^{*}\!+\!2$) than their brighter
counterparts. 

We find that the $B-R$ colour of a galaxy is dependent its environment,
whereas its luminosity is independent of local density, except at the very brightest magnitudes
\mbox{(M$_{R}<-22$)}. Red galaxies located on average in regions
with a mean local density greater than \mbox{1.0--1.1\,gals arcmin$^{-2}$} (\mbox{$\sim$200--250\,$h_{70}^{2}\,{\rm Mpc}^{-2}$}), while
blue galaxies are on average found in less dense regions. 

We find evidence for an increase in clustering of low-luminosity red
galaxies as also observed in the SDSS \citep{blanton04,zehavi}. The effect
appears to be greatest at \mbox{M$_{R}\sim-18$}, with fainter red galaxies found
in less dense environments. This is most likely to be related to the
dramatic effects of environment on the luminosity function of faint galaxies
observed for this region described in Paper I, in which dips are
apparent at \mbox{$-18\,\siml\!M_{R}\,\siml\!-$20}, which becomes deeper with
decreasing local density. The observed dips are due to a relative lack of red
sequence galaxies at these magnitudes in the lower density regions in
comparison to the cluster core, and so it is manifested here as an
enhanced clustering of red galaxies at  \mbox{M$_{R}\sim-18$}.
These observed changes in the galaxy luminosity function with
environment were explained in Paper I in
terms of the galaxy harassment scenario, in which infalling Sc--Sd spiral
galaxies are transformed by repeated high-velocity encounters over a
period of several Gyr, and in the process become 1--2\,mag dimmer due
to both mass loss and as a result of their star-formation being
terminated \citep*{moore98}. These galaxies which predominate in
\mbox{$z\sim0.4$} clusters at
\mbox{M$_{R}\sim-$20} are thus shifted to fainter
magnitudes, producing the observed dip, and hence the red galaxies
at \mbox{M$_{R}\sim-18$} are found preferentially in the highest-density
environments as they are produced as a direct result of harassment
by the supercluster environment. At fainter magnitudes the
harassment becomes too great, and in fact the dwarf galaxies are
completely disrupted by tidal forces in the cluster centre
\citep{moore98} and their stars cannibalised by the cD galaxy.

\subsection{Galaxy colours and the ICM}

The cores of each of the clusters are completely dominated by red
sequence galaxies, as indicated by regions containing very low fractions
\mbox{($<10$\%)} of blue galaxies, and red mean galaxy colours.
In the cases of A3562 and SC\,1329-313 these regions, which coincide with
both the centres of X-ray emission and the location of the brightest
cluster galaxy (BCG), are displaced from the peaks in the \mbox{$R<21$} galaxy
surface density, particularly for SC\,1329-313. The displacement, at
least between the X-ray emission/BCG and the peaks in the galaxy surface
density are likely to be related to the dynamics of the system, that
is an ongoing cluster merger, which has been proposed from an analysis
of the X-ray emission \citep{finoguenov} and the finding of a young
radio halo at the centre of A3562 \citep{venturi}. The strong
correlation between the regions devoid of blue galaxies and the
X-ray emission, as opposed to the density peaks, may be evidence in
support of ram pressure stripping having an important role in
terminating any remnant star-formation in galaxies that encounter the
cluster cores, although equally it could be a consequence of the
dynamics of the system displacing the faint galaxy population. 

The A3562--SC\,1329-313 system appears the most active in terms of
star-formation, with the highest densities of blue galaxies across the
supercluster survey region found here, in particular the
north-western flank of SC\,1329-313, but also between the two
clusters. These are also the regions where in a \mbox{7\,deg$^2$} VLA radio
survey of the whole supercluster core complex \citet{miller} found
dramatic increases in the fraction of \mbox{$\sim{\rm M}^{*}\!+1$} galaxies
belonging to the supercluster (i.e. with redshifts) which have
associated radio emission. From a visual inspection of their
morphology, these radio emitters are found to be spirals/irregulars,
and this region marks the greatest concentration of bright \mbox{$R<16$} spiral galaxies across the
supercluster complex. \citet{miller} interpreted these as young starburst
galaxies related to the cluster merger event, and taking this in
addition to our results which are sensitive to the dwarf galaxy
population, we can say this star-formation activity extends to \mbox{M$^{*}\!+6$}.

Are these galaxies which are undergoing starbursts triggered by the
effects of the cluster merger, for example by encountering shocks in
the ICM \citep{roettiger}, or are they are a group of normal
star-forming galaxies which have just entered the supercluster ?

Through an analysis of the X-ray emission from A3562 and SC\,1329-313,
\citet{finoguenov} argued that SC\,1329-313 had recently passed westwards
through the northern outskirts of A3562 and then been deflected south. This
would appear to contradict the idea that SC\,1329-313 is a remnant from the
ongoing A3562-A3558 merger, and instead suggests it represents a
recent arrival in the supercluster core complex.  
However the regions containing the largest concentrations of faint blue
galaxies are located {\em between} clusters, in particular along
the filament connecting the clusters A3558 and A3562, which would
support the hypothesis that shocks in the ICM produced by cluster
mergers is triggering starbursts in these galaxies.

\section{Conclusions}
\label{conclusions}
The area covered by the SOS dataset lies mostly within one
virial radius \mbox{(1--1.5\,$h_{70}$\,Mpc)} of one of the clusters.
The results presented here and in Paper I show that the global
properties of faint galaxies change significantly from the cluster
cores to the virial radius, both in terms of the luminosity function,
and the mean galaxy colours, which indicates that these galaxies are
being transformed by processes related to the supercluster
environment. As galaxy mergers should be infrequent in any of the
environments covered by the SOS, the finding of such large changes in
the mean galaxy colour or fraction of faint blue galaxies within the
SOS, indicates that some process other than merging must be
responsible. In paper I, we suggested that galaxy harassment is
important for shaping the galaxy luminosity function at magnitudes
fainter than \mbox{$\sim M^{*}+1$}, and here we find additional evidence in
favour of faint galaxies being transformed by ram-pressure stripping and
undergoing starbursts triggered by shocks in the ICM produced by cluster
mergers. 

These results indicate that the effect of environment on faint \mbox{($\ga
M^{*}+1$)} galaxies is quite different from that observed for bright
galaxies \citep{gomez,lewis}. While bright galaxies appear to be
transformed by processes that can take place well outside the virial
radius, that is galaxy merging or suffocation, we find here that many
faint galaxies are affected by their interaction with the
supercluster environment, although we cannot rule out that the
 faint galaxies are also transformed outside the virial radius, beyond the
 limits of our survey. It is also possible that the observed
 environmental trends are partly due to a population of primordial,
 early-type galaxies that formed preferentially in the high-density
 regions that later became clusters \citep[e.g.][]{poggianti05}.
 The differences in the environmental trends of bright and faint
 galaxies are likely to be related to the
observed global bimodality in galaxy properties
\citep[e.g.][]{kauffmann03} about a stellar characteristic stellar mass
of \mbox{$\sim3\times10^{10}\,{\rm M}_{\odot}$}. A possible
explanation could lie within the context of the hot and cold flow
model of galaxy evolution \citep{dekel,keres}. At low masses, galaxies
are able to merge and still retain their gas supply, and hence mergers
would have little effect on star-formation or galaxy colours. In
contrast, once galaxies merge to become more massive than a
characteristic mass, shocks in the halo become stable, and preventing
further cooling of gas from the halo, bringing about a transformation
in the galaxy star-forming properties.  

In a future article, we plan to compare our observational results with
predictions from semi-analytical models of galaxy evolution applied to
n-body simulations of a region containing a supercluster constrained
to match the dynamical structure of the Shapley supercluster region.

\section*{Acknowledgements}

We thank the anonymous referee for helpful comments that have improved
this paper.
CPH, AG and NK acknowledge the financial supports provided through
the European Community's Human Potential Program, under contract
HPRN-CT-2002-0031 SISCO. AM is supported by the Regione Campania
(L.R. 05/02) project {\em 'Evolution of Normal and Active Galaxies'}
and by the Italian Ministry of Education, University and Research
(MIUR) grant COFIN2004020323: {\em The Evolution of Stellar Systems: a 
Fundamental Step Towards the Scientific Exploitation of VST}. NK is
partially supported by the Italian Ministry of Education, University
and Research (MIUR) grant COFIN2003020150: {\em Evolution of Galaxies
  and Cosmic Structures after the Dark Age: Observational Study}.
We thank the Deep Lens Survey team and NOAO who provided the data used
to derive field galaxy counts.

\bsp

\label{lastpage}

\begin{thebibliography}{99}
\bibitem[\protect\citeauthoryear{Aarseth \& Fall}{1980}]{aarseth}
  Aarseth S.J., Fall S.M., 1980, ApJ, 236, 43
\bibitem[\protect\citeauthoryear{Abadi, Moore \& Bower}{Abadi et al.}{1999}]{a99}
  Abadi M.G., Moore B., Bower R.G., 1999, MNRAS, 308, 947
\bibitem[\protect\citeauthoryear{Babu \& Singh}{1983}]{babu}
  Babu G.J., Singh K., 1983, Ann. Stat., 11, 999
\bibitem[\protect\citeauthoryear{Baldry et al.}{2004}]{baldry}
  Baldry I.K., Glazebrook K., Brinkmann J., Ivezi\'{c}, \u{Z}., Lupton
  R.H., Nichol R.C., Szalay A.S., 2004, ApJ, 600, 681
\bibitem[\protect\citeauthoryear{Balogh et al.}{2004}]{balogh04}
  Balogh M.L., Baldry I.K., Nichol R., Miller C., Bower R., Glazebrook
  K., 2004, ApJL, 615, 101
\bibitem[\protect\citeauthoryear{Bardelli et al.}{1998}]{b98}
  Bardelli S., Zucca E., Zamorani G., Vettolani G., Scaramella R.,
  1998, MNRAS, 296, 599
\bibitem[\protect\citeauthoryear{Bardelli et al.}{2000}]{b00}
  Bardelli S., Zucca E., Zamorani G., Moscardini L., Scaramella R.,
  2000, MNRAS, 312, 540
\bibitem[\protect\citeauthoryear{Beers, Flynn \& Gebhardt}{Beers et al.}{1990}]{beers}
  Beers T.C., Flynn K., Gebhardt K., 1990, AJ, 100, 32
\bibitem[\protect\citeauthoryear{Bell et al.}{2004}]{bell}
  Bell E.F., Wolf C., Meisenheimer K., et al. 2004, ApJ, 608, 752 
\bibitem[\protect\citeauthoryear{Bertin \& Arnouts}{1996}]{bertin}
  Bertin E., Arnouts S., 1996, A\&AS, 331, 439
\bibitem[\protect\citeauthoryear{Blanton et al.}{2000}]{blanton00}
  Blanton M., Cen R., Ostriker J.P., Strauss M.A., Tegmark M., 2000,
  ApJ, 531, 1
\bibitem[\protect\citeauthoryear{Blanton et al.}{2003}]{blanton03}
  Blanton M., Hogg D.W., Bahcall N.A., et al., 2003, ApJ, 594, 186
\bibitem[\protect\citeauthoryear{Blanton et al.}{2004}]{blanton04}
  Blanton M., Eisenstein D., Hogg D.W., Zehavi I., 2004, ApJ in press (astro-ph/0411037)
\bibitem[\protect\citeauthoryear{Blanton et al.}{2005}]{blanton05}
  Blanton M., Eisenstein D., Hogg D.W., Schlegel D.J., Brinkmann J., 2005, ApJ, 629, 143
\bibitem[\protect\citeauthoryear{Bruzual \& Charlot}{2003}]{bruzual}
  Bruzual G., Charlot S., 2003, MNRAS, 344, 1000
\bibitem[\protect\citeauthoryear{Butcher \& Oemler}{1978}]{bo78}
  Butcher H., Oemler A., 1978, ApJ, 219, 18
\bibitem[\protect\citeauthoryear{Butcher \& Oemler}{1984}]{bo84}
  Butcher H., Oemler A., 1984, ApJ, 285, 426
\bibitem[\protect\citeauthoryear{Dekel \& Birnboim}{2006}]{dekel}
  Dekel A. Birnboim Y., 2006, MNRAS, 368, 2
\bibitem[\protect\citeauthoryear{Diaferio et al.}{2001}]{diaferio}
  Diaferio A., Kauffmann G., Balogh M.L., White S,D.M., Schade D.,
  Ellingson E., 2001, MNRAS, 232, 999
\bibitem[\protect\citeauthoryear{Dressler et al.}{1994}]{dressler94}
 Dressler A., Oemler A., Butcher H., Gunn J.E., 1994, ApJ, 430, 107
\bibitem[\protect\citeauthoryear{Dressler et al.}{1997}]{dressler97}
 Dressler A., et al., 1997, ApJ, 490, 577
\bibitem[\protect\citeauthoryear{Drinkwater et al.}{2004}]{drinkwater}
 Drinkwater M.J., Parker Q.A., Proust D., Slezak E. Quintana H., 2004,
 PASA, 21 89
\bibitem[\protect\citeauthoryear{Eisenstein et al.}{2003}]{eisenstein}
 Eisenstein D.J., Hogg D.W., Fukugita M., et al., 2003, ApJ, 585, 694
\bibitem[\protect\citeauthoryear{Faber et al.}{2005}]{faber}
 Faber S.M., Willmer C.N.A., Wolf C., et al., 2005, ApJ submitted, (astro-ph/0506044)
\bibitem[\protect\citeauthoryear{Finoguenov et al.}{2004}]{finoguenov}
 Finoguenov A., Henriksen M.J., Briel Y G., de Plaa J., Kaastra J.S.,
 2004, ApJ, 611, 811
\bibitem[\protect\citeauthoryear{Fujita \& Nagashima}{1999}]{fujita}
 Fujita Y., Nagashima M., 1999, ApJ, 516, 619
\bibitem[\protect\citeauthoryear{G\'{o}mez et al.}{2003}]{gomez}
G\'{o}mez P.L., Nichol R.C., Miller C.J., et al., 2003, ApJ, 584, 210
\bibitem[\protect\citeauthoryear{Ghigna et al.}{1998}]{ghigna}
Ghigna S., Moore B., Governato F., Lake G., Quinn T., Stadel J., 1998,
MNRAS, 300, 146
\bibitem[\protect\citeauthoryear{Gray et al.}{2004}]{gray}
  Gray M.E, Wolf C., Meisenheimer K., Taylor A., Dye S., Borch A.,
  Kleinheinrich M., 2004, MNRAS, 347, L73
\bibitem[\protect\citeauthoryear{Haines et al.}{2004}]{haines04}
  Haines C.P., Mercurio A., Merluzzi P., La Barbera F., Massarotti M.,
  Busarello G., Girardi M., 2004, A\&A, 425, 783
\bibitem[\protect\citeauthoryear{Hogg et al.}{2004}]{hogg}
  Hogg D.W., Blanton M.R., Brinchmann J., et al., 2004, ApJL, 601, 29 
\bibitem[\protect\citeauthoryear{Kauffmann}{1995}]{kauffmann95}
 Kauffmann G., 1995, MNRAS, 274, 153
\bibitem[\protect\citeauthoryear{Kauffmann et al.}{2003}]{kauffmann03}
 Kauffmann G., Heckman T.M., White S.D.M., et al., 2003, MNRAS, 341, 54
\bibitem[\protect\citeauthoryear{Kere\v{s} et al.}{2005}]{keres}
 Kere\v{s} D., Katz, N., Weinberg, D.H., Dav\'{e}, R. 2005, MNRAS, 363, 2 
\bibitem[\protect\citeauthoryear{Kodama \& Bower}{2001}]{kodama01}
Kodama T.,  Bower R.G., 2001, MNRAS, 321, 18
\bibitem[\protect\citeauthoryear{Kodama et al.}{2004}]{kodama}
Kodama T., Balogh M.L., Smail I., Bower R.G., Nakata F., 2004, MNRAS,
 354, 1103
\bibitem[\protect\citeauthoryear{Landolt}{1992}]{landolt}
 Landolt A.U., 1992, AJ, 104, 340
\bibitem[\protect\citeauthoryear{Larson, Tinsley \& Caldwell}{Larson et al.}{1980}]{larson}
 Larson R.B., Tinsley B.M., Caldwell C.N., 1980, 237, 692
\bibitem[\protect\citeauthoryear{Lewis et al.}{2002}]{lewis}
 Lewis I., Balogh M., de Propris R., et al., 2002, MNRAS, 334, 673 
\bibitem[\protect\citeauthoryear{L\'{o}pez-Cruz, Barkhouse \& Yee}{L\'{o}pez-Cruz et al.}{2004}]{lopez}
 L\'{o}pez-Cruz O., Barkhouse, W.A., Yee H.K.C., 2004, ApJ, 614,679
\bibitem[\protect\citeauthoryear{Mercurio et al.}{2006}]{paper1}
 Mercurio A., Merluzzi P., Haines C.P., Gargiulo A., Krusanova N.,
 Busarello G., La Barbera F., 2006,  MNRAS, 368, 109
\bibitem[\protect\citeauthoryear{Metcalfe, Godwin \& Peach}{Metcalfe et al.}{1994}]{metcalfe}
 Metcalfe N., Godwin J.G., Peach J.V., 1994, MNRAS, 267, 431
\bibitem[\protect\citeauthoryear{Miller}{2005}]{miller}
 Miller N.A., 2005, AJ, 130, 2541
\bibitem[\protect\citeauthoryear{Moore et al.}{1996}]{moore}
 Moore B., Katz N., Lake G., Dressler A., Oemler A. Jr., 1996, Nature,
 379, 613
\bibitem[\protect\citeauthoryear{Moore, Lake \& Katz}{Moore et al.}{1998}]{moore98}
 Moore B., Lake G., Katz N., 1998, ApJ, 495, 139 
\bibitem[\protect\citeauthoryear{Pisani}{1993}]{pisani93}
 Pisani A., 1993, MNRAS, 265, 706
\bibitem[\protect\citeauthoryear{Pisani}{1996}]{pisani96}
 Pisani A., 1996, MNRAS, 278, 697
\bibitem[\protect\citeauthoryear{Poggianti}{1997}]{poggianti}
 Poggianti B., 1997, A\&AS, 122, 399
\bibitem[\protect\citeauthoryear{Poggianti}{2006}]{poggianti05}
 Poggianti B., von der Linden, A., de Lucia, G., et al. 2006, ApJ,
 642, 188
\bibitem[\protect\citeauthoryear{Quintana, Carrasco \& Reisenegger}{Quintana et al.}{2000}]{quintana}
 Quintana H., Carrasco H., Reisenegger A., 2000, AJ, 120, 511
\bibitem[\protect\citeauthoryear{Reisenegger et al.}{2000}]{reisenegger}
 Reisenegger A., Quintana H., Carrasco H., Maze, J. 2000, AJ, 120, 523
\bibitem[\protect\citeauthoryear{Roettiger, Burns \& Loken}{Roettiger et al.}{1996}]{roettiger}
 Roettiger K., Burns J.O., Loken C., 1996, ApJ, 473, 651 
\bibitem[\protect\citeauthoryear{Shapley}{1930}]{shapley}
 Shapley H., 1930, Bull. Harvard Obs., 874, 9
\bibitem[\protect\citeauthoryear{Strateva et al.}{2001}]{strateva}
 Strateva I., Ivez\'{i}c \u{Z}., Knapp G.R. et al., 2001, AJ, 122, 1861  
\bibitem[\protect\citeauthoryear{Tanaka et al.}{2004}]{tanaka}
  Tanaka M., Goto T., Okamura S., Shimasaku K., Brinkmann J., 2004,
  AJ, 128, 2677
\bibitem[\protect\citeauthoryear{Thomas et al.}{2005}]{thomas}
  Thomas D., Maraston C., Bender R., Mendes de Oliveira C., 2005, ApJ,
  621, 673
\bibitem[\protect\citeauthoryear{Treu et al.}{2003}]{treu}
  Treu T., Ellis R., Kneib J.-P., Dressler A., Smail I., Czoske O.,
  Oemler A., Natarajan R., 2003, ApJ, 591, 53
\bibitem[\protect\citeauthoryear{Tremonti et al.}{2004}]{tremonti}
  Tremonti, C.A., Heckman, T.M., Kauffmann G., et al. 2004, ApJ, 613, 898
\bibitem[\protect\citeauthoryear{Vandame}{2004}]{vandame}
  Vandame, 2004, PhD thesis
\bibitem[\protect\citeauthoryear{Venturi et al.}{2000}]{venturi}
  Venturi T., Bardelli S., Morganti R., Hunstead R.W., 2000, MNRAS,
  314, 594
\bibitem[\protect\citeauthoryear{Wittman et al.}{2002}]{wittman}
 Wittman B.M., Tyson J.A., Dell'Antonio I.P. et al., 2002, SPIE, 4836, 73
\bibitem[\protect\citeauthoryear{Zehavi et al.}{2005}]{zehavi}
 Zehavi I., Zheng Z., Weinberg D.H., et al., 2005, ApJ, 630, 1 

\end{thebibliography}
\end{document}